\documentclass[11pt]{article}
\usepackage[noblocks]{authblk}
\usepackage{titlesec}
\usepackage{geometry}
\usepackage{multirow}
\usepackage[square,sort,comma,numbers]{natbib}
\usepackage{graphicx}
\usepackage[english]{babel}
\usepackage{amsmath}
\usepackage{amssymb}
\usepackage{amsfonts}
\usepackage{amsthm}
\usepackage{thmtools} 
\usepackage{thm-restate}
\usepackage{accents}
\usepackage[noend]{algpseudocode}
\usepackage{caption}
\usepackage{subcaption}

\makeatletter
\let\OldStatex\Statex
\renewcommand{\Statex}[1][3]{
  \setlength\@tempdima{\algorithmicindent}
  \OldStatex\hskip\dimexpr#1\@tempdima\relax}
\makeatother

\algnewcommand{\Inp}{\textbf{Input:}\space}
\algnewcommand{\Out}{\textbf{Output:}\space}
\newcommand{\Input}{\Statex[-1] \Inp }
\newcommand{\Output}{\Statex[-1] \Out }
\newcommand{\Blank}{\Statex[-1]}

\usepackage[dvipsnames]{xcolor}
\definecolor{cb-black}      {RGB}{  0,   0,   0}
\definecolor{cb-blue-green} {RGB}{  0,  073,  073}
\definecolor{cb-green-sea}  {RGB}{  0, 146, 146}
\definecolor{cb-rose}       {RGB}{255, 109, 182}
\definecolor{cb-salmon-pink}{RGB}{255, 182, 119}
\definecolor{cb-purple}     {RGB}{ 73,   0, 146}
\definecolor{cb-blue}       {RGB}{ 0, 109, 219}
\definecolor{cb-lilac}      {RGB}{182, 109, 255}
\definecolor{cb-blue-sky}   {RGB}{109, 182, 255}
\definecolor{cb-blue-light} {RGB}{182, 219, 255}
\definecolor{cb-burgundy}   {RGB}{146,   0,   0}
\definecolor{cb-brown}      {RGB}{146,  73,   0}
\definecolor{cb-clay}       {RGB}{219, 209,   0}
\definecolor{cb-green-lime} {RGB}{ 36, 255,  36}
\definecolor{cb-yellow}     {RGB}{255, 255, 109}

\usepackage[colorlinks,linkcolor=cb-burgundy,urlcolor=cb-burgundy,citecolor=cb-burgundy,destlabel]{hyperref}
\usepackage[nameinlink,capitalise]{cleveref}
\usepackage{nicefrac}
\usepackage{stmaryrd}
\usepackage[normalem]{ulem}

\makeatletter
\providecommand\theHALG@line{\thealgorithm.\arabic{ALG@line}}
\makeatother

\newtheorem{theorem}{Theorem}
\counterwithin{figure}{section}
\counterwithin{table}{section}
\counterwithin{theorem}{section}

\theoremstyle{remark}

\declaretheoremstyle[
  notefont=\mdseries, notebraces={(}{)},
  bodyfont=\normalfont,
  postheadspace=0em,
  headpunct=
]{algostyle}
\theoremstyle{algostyle}

\newtheorem{algorithm}[theorem]{Algorithm}

\newcommand*{\f}{\mathbb{F}}

\newcommand*{\ip}[1]{ \langle #1 \rangle }

\makeatletter
\def\smalloverbrace#1{\mathop{\vbox{\m@th\ialign{##\crcr\noalign{\kern3\p@}
  \tiny\downbracefill\crcr\noalign{\kern3\p@\nointerlineskip}
  $\hfil\displaystyle{#1}\hfil$\crcr}}}\limits}
\makeatother

\title{Clifford synthesis via generalized S and CZ gates}

\author[1]{Vadym Kliuchnikov\thanks{Current address: NVIDIA, Toronto, ON M5V 1K4, Canada}}
\affil[1]{Microsoft Quantum, Toronto, ON M5J 0E7, Canada}
\author[2]{Marcus P. da Silva}
\affil[2]{Microsoft Quantum, Redmond, WA 98052, USA}

\begin{document}

\maketitle

\begin{abstract}
We show that any $n$-qubit Clifford unitary can be implemented using at most $2n$ 
multi-qubit joint measurements.
All the multi-qubit joint measurements used for implementing the Clifford unitary 
can be chosen to form at most two sets of 
independent mutually-commuting measurements.
Each of these sets is of size at most $n$.
This enables very flexible space-time trade-offs when implementing Clifford unitaries.
We also discuss a version of the result that relies on multi-target CNOTs and is more relevant for targeting fault-tolerant hardware 
based on Quantum LDPC codes.
\end{abstract}

\section{Introduction}

We describe an algorithm for synthesizing an $n$-qubit Clifford unitary using at most $2n$ multi-qubit joint measurements.
Moreover, all the multi-qubit joint measurements used for implementing the Clifford unitary 
can be chosen to form at most two sets of independent mutually-commuting measurements.
Each of these sets is of size at most $n$.
This is useful when trying to implement an arbitrary $n$-qubit Clifford unitary using lattice-surgery surface codes.
This problem naturally comes up when switching between different layout methods for mapping a quantum algorithm to fault-tolerant hardware,
in particular when switching from Pauli-Based Computation~\cite{GSC2019, aasen2025topologicallyfaulttolerantquantumcomputer}
or Parallel Synthesis Sequential Pauli Computation~\cite{B2022} layout methods to other layout methods, such as Edge Disjoint Paths~\cite{EDP2022}.
Having two sets of mutually-commuting joint measurements enables more flexibility for space-time trade-offs.

\section{Preliminaries}

Here we review some well-known results that we use in our Clifford synthesis algorithm.

\subsection{Pauli unitaries}

Recall that one-qubit Pauli matrices are
$$
I = 
\left(
\begin{array}{cc}
    1 & 0 \\
    0 & 1
\end{array}
\right),\,
X = 
\left(
\begin{array}{cc}
    0 & 1 \\
    1 & 0
\end{array}
\right),\,
Y =
\left(
\begin{array}{cc}
    0 & -i \\
    i & 0
\end{array}
\right),\,
Z =
\left(
\begin{array}{cc}
    1 & 0 \\
    0 & -1
\end{array}
\right),\,
$$
$n$-qubit Pauli matrices are $\{I,X,Y,Z\}^{\otimes n}$, the $n$-fold tensor products of one-qubit Pauli matrices.
Pauli observables are $n$-qubit Hermitian matrices $\pm \{I,X,Y,Z\}^{\otimes n}$. 
For one-qubit Pauli observables $P$, we define x bits $x(P)$ as: 
$$
x(\pm I) = 0,\,x(\pm X)=1,\,x( \pm Y)=1,\,x(\pm Z)=0,
$$
and z bits $z(P)$ as:
$$
z(\pm I) = 0,\,z(\pm X)=0,\,z(\pm Y)=1,\,z(\pm Z)=1.
$$
For an $n$-qubit Pauli observable $P = \pm P_1 \otimes \ldots \otimes P_n $ we define x and z bits as:
$$
    x(P) = (x(P_1),\ldots,x(P_n)),\quad z(P) = (z(P_1),\ldots,z(P_n)).
$$
It is well known that the group commutator of two Pauli observables $P$, $Q$ can be written as 
$$
 \llbracket P,Q \rrbracket = PQP^\dagger Q^\dagger = (-1)^{\ip{x(P),z(Q)} + \ip{z(P),x(Q)} } (I\otimes \ldots \otimes I),
$$
where $\ip{x(P),z(Q)}$ is the inner product of vectors $x(P)$, $z(Q)$ with addition and multiplication modulo $2$,
that is $1 + 1 = 0$.

It is common to use $X_j,Z_j$ for an $n$-qubit Pauli observable acting as $X,Z$ on qubit $j$ and as $I$ on the rest of the qubits.

Finally, recall that the group commutator of two Pauli observables is related to a symplectic inner product.
We rewrite the sum $\ip{x(P),z(Q)} + \ip{z(P),x(Q)}$ as 
an inner product of vectors $z(P) \oplus x(P)$ and $z(Q) \oplus x(Q)$, where $\oplus$ denotes the direct sum of two vectors.
Define the $2n \times 2n$ matrix $\Omega$ as
\begin{equation}
\label{eq:omega}
 \Omega =
\left(
\begin{array}{c|c}
    0 & I_n \\ \hline
    I_n & 0
\end{array}
\right),
\end{equation}
where $I_n$ is the $n\times n$ identity matrix. Using matrix $\Omega$ we have:
$$
\ip{x(P),z(Q)} + \ip{z(P),x(Q)} = \ip{ \Omega (z(P)\oplus x(P)), z(Q)\oplus x(Q) }.
$$
Again, all addition and multiplication operations are modulo two.
The map $x,y \mapsto \ip{ \Omega x,y}$ is an example of a symplectic inner product.

\subsection{Clifford unitaries}

Recall that a unitary $C$ is a Clifford unitary if it maps Pauli observables to 
Pauli observables by conjugation, that is, for any Pauli observable $P$
we have that $CPC^\dagger$ is also a Pauli observable.
It is common to associate with a Clifford unitary $C$ a $2n\times 2n$ matrix with $\{0,1\}$-entries:
\begin{equation}
\label{eq:binary-symplectic-matrix-definition}
M_C = 
\left(
\begin{array}{c|c}
    A_{x,x} & A_{x,z} \\ \hline
    A_{z,x} & A_{z,z}
\end{array}
\right),
\end{equation}
where the rows of matrices $A_{z,z},  A_{z,x},  A_{x,z},  A_{x,x}$ are defined 
by the following equations 
\begin{align*}
    (A_{z,z})_j = & z(CZ_jC^\dagger),& (A_{z,x})_j = & x(CZ_jC^\dagger),& j \in& \{1,\ldots,n\} \\
    (A_{x,z})_j = & z(CX_jC^\dagger),& (A_{x,x})_j = & x(XZ_jC^\dagger),& j \in& \{1,\ldots,n\} 
\end{align*}

Recall that conjugation by a Clifford unitary preserves the group commutator; therefore: 
\begin{align*}
 \llbracket CZ_jC^\dagger, CX_jC^\dagger \rrbracket = & \llbracket Z_j, X_k \rrbracket, \\
 \llbracket CZ_jC^\dagger, CZ_jC^\dagger \rrbracket = & \llbracket Z_j, Z_k \rrbracket, \\
 \llbracket CX_jC^\dagger, CX_jC^\dagger \rrbracket = & \llbracket X_j, X_k \rrbracket.
\end{align*}
The relation between the group commutator of Pauli observables and the symplectic inner product implies 
that $M_C$ is a symplectic matrix, that is, $M_C^T \Omega M_C = \Omega$, where $\Omega$ is defined in \cref{eq:omega}.
Note that the matrix $M_C$ defines the Clifford unitary $C$ up to multiplication by a Pauli matrix $P$ 
and a global phase $e^{i \phi}$; that is, for unitaries $C$ and $D$, equality $M_C = M_D$ implies that
there exist a Pauli matrix $P$ and a real number $\phi$ such that 
$
 C = e^{i\phi} D P.
$
Additionally, $M_{C D} = M_C M_D$ and any symplectic matrix corresponds to a Clifford unitary~\cite{AaronsonGottesman2004}.
For these reasons, questions about synthesis of Clifford unitaries 
are closely related to results on symplectic matrices with entries in the field $\f_2$, that is, the field with elements $\{0,1\}$
and addition and multiplication modulo two.

Recall some common examples of Clifford unitaries.
The $S$ unitary is a one-qubit matrix $e^{-i\pi Z/4 }$, and the $CZ$ unitary is a two-qubit matrix $e^{i \pi (I-Z)\otimes (I-Z)/4}$.
The corresponding symplectic matrices are 
\begin{equation} \label{eq:symplectic-examples}
     M_S = \left(
\begin{array}{c|c}
    1 & 1 \\ \hline
      & 1
\end{array}
\right),\quad 
M_{CZ} = 
\left(
\begin{array}{cc|cc}
    1 &   & 0 & 1 \\ 
      & 1 & 1 & 0 \\ \hline
      &   & 1 &   \\
      &   &   & 1 
\end{array}
\right).
\end{equation}
The generalized $S$ gate is defined for a Pauli observable $P$ as $e^{i \pi P/4}$.
The generalized $CZ$ gate is defined for two commuting independent Pauli observables $P$, $Q$ as 
$\Lambda(P,Q) = e^{ i\pi (I-P)(I-Q)/4}$~\cite{GSC2019}. 
Conjugating $S$ acting on the first qubit and $CZ$ acting on the first two qubits by a Clifford unitary $C$ maps them to generalized $S$ and $CZ$
gates with $P = CZ_1 C^\dagger, Q = C Z_2 C^\dagger$.

A Pauli unitary on $n$ qubits is represented as a sequence of $n$ letters $I,X,Y,Z$ corresponding to the four one-qubit Pauli matrices, so that 
$$
P = P_1 \otimes \ldots \otimes P_n, P_k \in \{ I,X,Y,Z \}.
$$

Any $n$-qubit Clifford unitary $C$ can be fully described by a $(2n+1)\times(2n+1)$ binary matrix as discussed in~\cite{DDM2003}. 
The binary symplectic matrix $M_C$ in \cref{eq:cz-product} corresponds to the top-left $2n\times2n$ sub-matrix of the 
$(2n+1)\times(2n+1)$ matrix describing the Clifford unitary $C$ completely. 
When Clifford unitaries are represented by $(2n+1)\times(2n+1)$ matrices, their product and inverse (denoted by $\dagger$) can be computed efficiently~\cite{DDM2003}.
A Clifford unitary $C$ for which $M_C$ is the identity is a Pauli unitary. 
Which Pauli unitary $P$ corresponds to such a Clifford
can be efficiently computed by using the $(2n+1)\times(2n+1)$ binary matrix describing the Clifford~\cite{DDM2003}.

When we say that Clifford unitary $B_1$ is such that $M_{B_1} = F_1$ for some binary-symplectic matrix $F_1$, it means that $B_1$ is represented by a $(2n+1)\times(2n+1)$
binary matrix with the top-left $2n\times 2n$ part equal to $F_1$ and the rest of the entries chosen to be consistent with the requirements of storing a Clifford unitary as a
$(2n+1)\times(2n+1)$ matrix.

\subsection{Symplectic matrices}
Recall that a symplectic matrix over $\f_2$ is a $2n\times2n$ matrix $M$ such that $M^T \Omega M = \Omega$ where 
$\Omega$ is defined in \cref{eq:omega}.
A symplectic matrix $M$ is a symplectic involution if $M^2 = I_{2n}$, that is, the square of $M$ is the $2n\times 2n$ identity matrix.

There are two notable results on involutions of symplectic matrices over $\f_2$. The first one is:
\begin{theorem}[The main theorem in \cite{GOW1981}] \label{thm:two-involutions}
Each symplectic matrix is a product of two symplectic involutions.
\end{theorem}
This result implies that any question about finding a circuit for a Clifford unitary $C$
can be reduced to a question about finding circuits for Clifford unitaries $B,D$
such that $M_C = M_B M_D$ and $M_B^2 = I$, $M_D^2 = I$.
The next notable result concerns the structure of symplectic involutions.
\begin{theorem}[Theorem~2.1.16 in \cite{O1978}] \label{thm:symplectic-involution-basis}
Let $R$ be a $2n\times2n$ symplectic involution. Then there exists a symplectic matrix $M$ such that 
\begin{equation} \label{eq:involution-structure}
M^{-1} R M = \left(
\begin{array}{c|c}
    I & A \\ \hline
      & I
\end{array}
\right),
A=
\left(
\begin{array}{c|c}
    A' &  \\ \hline
     & 0
\end{array}
\right)
\end{equation}
where $A'$ is an identity matrix, or $A'$ is a block-diagonal matrix:
\begin{equation} \label{eq:involution-block-diagonal}
A'=
\left(
\begin{array}{ccc}
   \begin{array}{cc} 0 & 1 \\ 1 & 0 \end{array} & & \\ 
     & \ddots & \\
     & & \begin{array}{cc} 0 & 1 \\ 1 & 0 \end{array}
\end{array}
\right)\end{equation}
\end{theorem}
Matrices $M_S$, $M_{CZ}$ from \cref{eq:symplectic-examples} are involutions and are
in the form given by \cref{eq:involution-structure,eq:involution-block-diagonal} with $M = I$.
An immediate corollary of the theorem is that for any $n$-qubit Clifford unitary $B$ such that 
$M_B$ is an involution, there exists a Clifford unitary $C$ such that one of the following holds: 
\begin{align}
\text{(A) there exist } & k \le n, \text{ Pauli } P, \text{ real number } \phi :& & B = e^{i \phi} C \left( S^{\otimes k} \otimes I^{\otimes n-k} \right) C^\dagger P \label{eq:s-product} \\
\text{(B) there exist } & k \le n/2, \text{ Pauli } P, \text{ real number } \phi :& & B = e^{i \phi} C \left( CZ^{\otimes k} \otimes I^{\otimes n-2k} \right) C^\dagger P \label{eq:cz-product}.
\end{align}
with the alternative (A) corresponding to the case where $A'$ is a $k\times k$ identity matrix and alternative (B) corresponding to the case when $A'$ is a $2k\times 2k$ block-diagonal matrix. 
In other words, any Clifford unitary $B$ with $M_B$ being a symplectic involution is a product of either generalized $S$ or generalized $CZ$ gates.

\subsection{Remote execution of diagonal gates and circuits for generalized S and CZ gates}

\begin{figure}[p]
    \centering
    \includegraphics[scale=0.5]{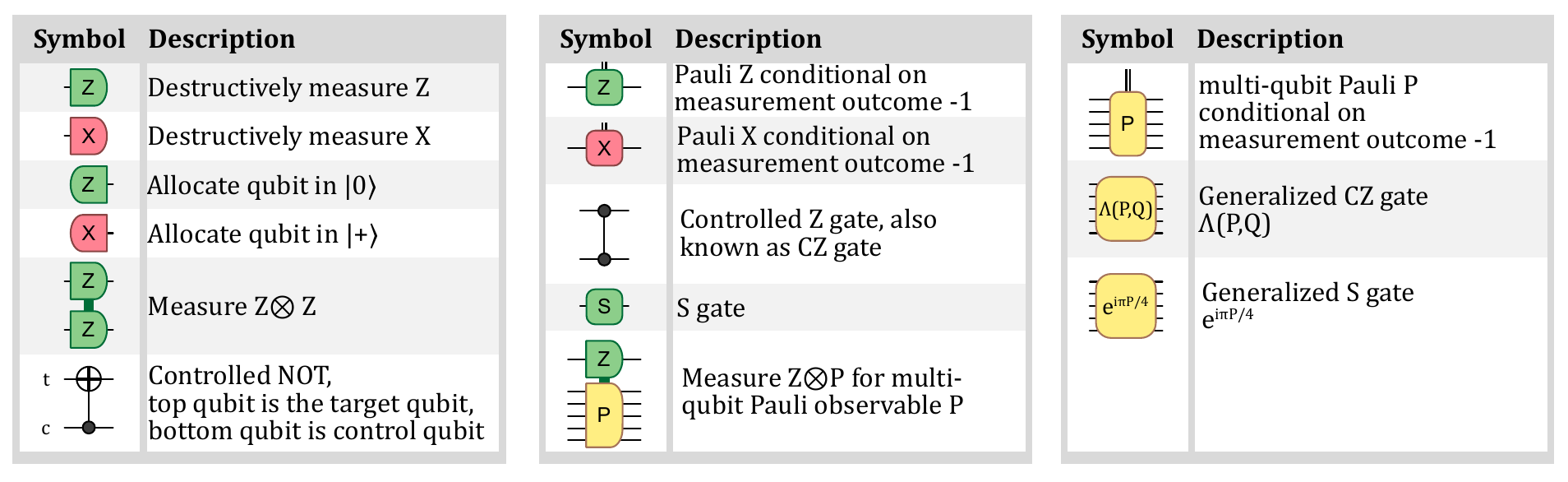}
    \caption{Quantum circuit notation}
    \label{fig:legend}
\end{figure}

\begin{figure}[p]
    \centering
    \includegraphics[scale=0.6]{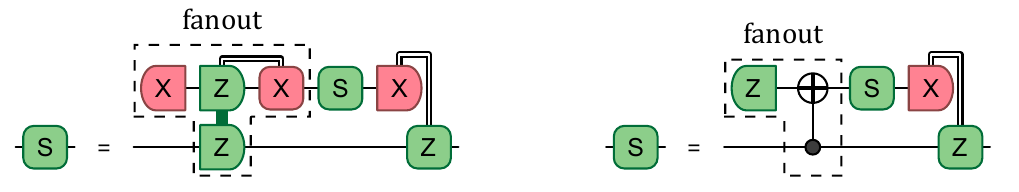}
    \caption{Remote execution of S gate. The circuit on the right is described in Appendix D in~\cite{EDP2022}; the circuit on the left is obtained from the circuit on the right by replacing the CNOT-based fanout gate with 
    a measurement-based one, similar to~\cite{B2022}.}
    \label{fig:remote-s}
\end{figure}

\begin{figure}[p]
    \centering
    \includegraphics[scale=0.6]{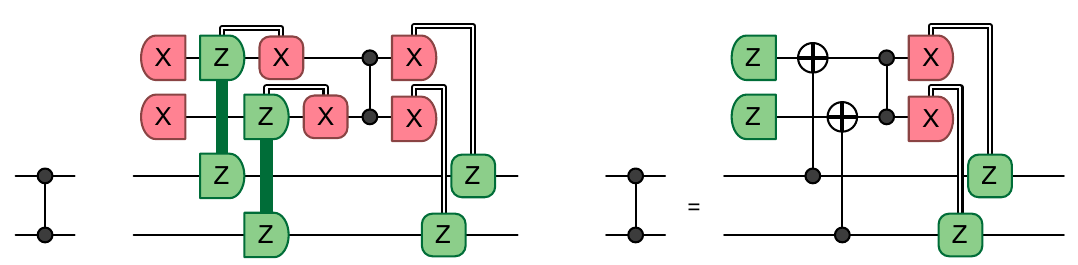}
    \caption{Remote execution of CZ gate. The circuit on the right is described in Appendix D in~\cite{EDP2022}; the circuit on the left is obtained from the circuit on the right by replacing the CNOT-based fanout gate with 
    a measurement-based one, similar to~\cite{B2022}.}
    \label{fig:remote-cz}
\end{figure}

We use quantum circuit diagram notation in \cref{fig:legend}.
Let us recall the circuits~\cref{fig:remote-s,fig:remote-cz} for remote execution of diagonal gates from Appendix~D in~\cite{EDP2022} that lead to 
circuits for execution of generalized $S$ gates~(similar to \cite{GSC2019}) and generalized $CZ$ gates~(similar to executing $CCZ$ during Parallel Synthesis Sequential Pauli Computation in \cite{B2022}).

\begin{figure}[p]
    \centering
    \includegraphics[scale=0.6]{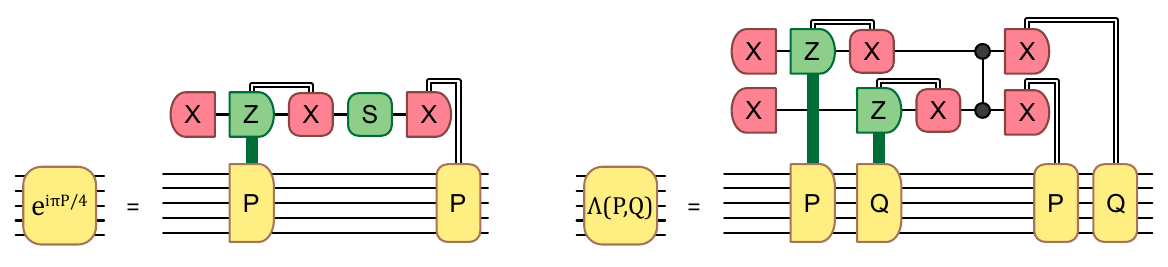}
    \caption{Circuits for executing generalized S and CZ gates using one and two multi-qubit Pauli measurements. Similar circuits are in Figure~11(b) and Figure~33(b) in~\cite{GSC2019}.}
    \label{fig:remote-gen-s-cz}
\end{figure}

Recall that for any commuting independent Pauli observables $P$, $Q$ there exists a Clifford unitary $C$ such that $CZ_1C^\dagger = P$ and $CZ_2C^\dagger = Q$.
Conjugating the bottom qubit in the left circuit in~\cref{fig:remote-s} by such $C$ leads to the circuit for the generalized S gate in~\cref{fig:remote-gen-s-cz}, similar to Figure~11(b) in~\cite{GSC2019}.
Conjugating the bottom two qubits in the left circuit in~\cref{fig:remote-cz} by such $C$ leads to the circuit for the generalized CZ gate in~\cref{fig:remote-gen-s-cz},
similar to Figure~33(b) in~\cite{GSC2019}.
If one is interested in using generalized CZ gates $\Lambda(X_1,P)$, $\Lambda(X_2,Q)$ as in~\cite{YR2023} instead of multi-qubit joint measurements $Z_1 P$, $Z_2 Q$ in~\cref{fig:remote-gen-s-cz},
it is sufficient to conjugate the bottom half of the qubits in the right circuits in~\cref{fig:remote-s,fig:remote-cz}.

\section{Clifford synthesis algorithm}

We combine results from the previous section into a Clifford synthesis algorithm:

\begin{algorithm}[Clifford synthesis via involutions]
\begin{algorithmic}[1]
\Blank
\Input An $n$-qubit Clifford unitary $C$
\Output Sequences $s_1$, $s_2$ of generalized $S$ and $CZ$ gates, and a Pauli unitary $P$, such that
\begin{itemize}
    \item Gates from sequences $s_1$ and $s_2$ followed by Pauli $P$ implement $C$
    \item Generalized $S$ and $CZ$ gates from $s_1$ are implemented using mutually-commuting multi-qubit measurements~(\cref{fig:remote-gen-s-cz})
    \item Generalized $S$ and $CZ$ gates from $s_2$ are implemented using mutually-commuting multi-qubit measurements~(\cref{fig:remote-gen-s-cz})
    \item Sequences $s_1$, $s_2$ both require at most $n$ multi-qubit measurements when using circuits in~(\cref{fig:remote-gen-s-cz})
\end{itemize}
\State Decompose $M_C$ into a product of two involutions $M_1 M_2$ using \cref{thm:two-involutions}.
\State Find symplectic matrices $F_1,F_2$ that bring $M_1$, $M_2$ into a special form with matrices $A_1',A_2'$ using \cref{thm:symplectic-involution-basis}.
\State Let $B_1, B_2$ be Clifford unitaries such that $M_{B_1} = F_1$, $M_{B_2} = F_2$.
\If{$A_1'$ is a diagonal matrix}
\State Set $k$ to the size of $A_1'$
\State Set $s_2$ to the sequence $e^{i\pi B_1 Z_1 B^\dagger_1}, \ldots, e^{i\pi B_1 Z_k B^\dagger_1}$
\Else \Comment $A_1'$ is a block-diagonal matrix
\State Set $k$ to half of the size of $A_1'$ 
\State Set $s_2$ to the sequence $\Lambda(B_1 Z_1 B^\dagger_1, B_1 Z_2 B^\dagger_1),\ldots,\Lambda(B_1 Z_{2k-1} B^\dagger_1, B_1 Z_{2k} B^\dagger_1)$ 
\EndIf
\If{$A_2'$ is a diagonal matrix}
\State Set $k$ to the size of $A_2'$
\State Set $s_1$ to the sequence $e^{i\pi B_2 Z_2 B^\dagger_1}, \ldots, e^{i\pi B_2 Z_k B^\dagger_2}$
\Else \Comment $A_2'$ is a block-diagonal matrix
\State Set $k$ to half of the size of $A_2'$ 
\State Set $s_1$ to the sequence $\Lambda(B_2 Z_1 B^\dagger_2, B_2 Z_2 B^\dagger_2),\ldots,\Lambda(B_2 Z_{2k-1} B^\dagger_2, B_2 Z_{2k} B^\dagger_2)$ 
\EndIf
\State Set $P \leftarrow \left(\prod_{g \in s_2} g \right)\left( \prod_{g \in s_1} g \right) C^\dagger$  
\State \Return $s_1,s_2,P$
\end{algorithmic}
\end{algorithm}

We conclude with a brief discussion of the algorithm correctness. 
Let $C_1 = \left( \prod_{g \in s_1} g \right)$, $C_2 = \left( \prod_{g \in s_2} g \right)$ be the Clifford unitaries 
implemented by sequences of gates $s_1, s_2$.
It must be that $M_{C_1} = M_1$, and $M_{C_2} = M_2$, and consequently $M_C = M_{C_1} M_{C_2}$.
Indeed, symplectic matrices of products $e^{i\pi  Z_1  }, \ldots, e^{i\pi  Z_k }$
or $\Lambda( Z_1, Z_2),\ldots,\Lambda(Z_{2k-1}, Z_{2k})$, correspond to $F_1^{-1} M_1 F_1$.
Conjugating products $e^{i\pi  Z_1  }, \ldots, e^{i\pi  Z_k }$
or $\Lambda( Z_1, Z_2),\ldots,\Lambda(Z_{2k-1}, Z_{2k})$ by $B_1$ gives us $C_1$ and $s_1$ and transforms 
$F_1^{-1} M_1 F_1$ into $M_1$. 
Similarly, we see that $M_{C_2} = M_2$.
Each of the sequences $s_1$, $s_2$ is implemented using joint measurements $Z\otimes B_1 Z_j B_1^\dagger$, $Z\otimes B_2 Z_j B_2^\dagger$,
which form two sets of mutually-commuting measurements, as required.
Finally, the number of measurements in each group is equal to the size of $A_1'$, $A_2'$ and is at most $n$.

\section{Clifford synthesis using A* search}
\label{sec:a-star}

We also note that the $A^\ast$ graph search algorithm can be used to synthesize a Clifford unitary using generalized $S$ and $CZ$ gates,
with the cost function being the number of Pauli measurements used, similar to its use for other gate-sets in \cite{webster2025heuristicoptimalsynthesiscnot}. 
The vertices of the graph are symplectic matrices $M_C$ of Clifford unitaries.
Two vertices $M_{C_1}$, $M_{C_2}$ are connected by an edge with weight one if $C_1 = e^{i \pi P/4} C_2 $,
and are connected by an edge with weight two if $C_1 = \Lambda(P,Q) C_2 $.

It remains to define a heuristic function $h(M_C)$ that gives an estimated cost from current node $M_C$ to goal $I$.
We recall another well-known integer-valued function $\mathrm{res}$~\cite{O1978} associated with a $2n\times 2n$ symplectic matrix $M$:
$$
 \mathrm{res}(M) = \mathrm{dim}((M-I_{2n})\f_2^{2n})
$$
Notably, for generalized $S$ and $CZ$ gates we have 
$$
 \mathrm{res}(M_{e^{i\pi P/4}}) = 1,\, \mathrm{res}(M_{\Lambda(P,Q)}) = 2.
$$
That is, $\mathrm{res}$ of the corresponding symplectic matrix is equal to the number of Pauli measurements required to implement generalized $S$
and $CZ$ gates.

We use search heuristic function $h(M_C) = \mathrm{res}(M_C)$.
Next we show that $h(M_C)$ is an admissible heuristic, that is, it provides a lower bound on the true cost of reaching the goal.
This implies that applying A* search to synthesizing $C$ will lead to a sequence of generalized $S$ and $CZ$ gates that uses the smallest possible 
number of Pauli measurements.
Suppose now that $s(C)$ is a sequence of generalized $S$ and $CZ$ gates that implements Clifford $C$ and uses the smallest possible 
number of Pauli measurements $m(C)$. 
Using the multiplicative property of $\mathrm{res}$ from Theorem~1.3.2 in~\cite{O1978}: 
$$
 \mathrm{res}(M_1 M_2) \le  \mathrm{res}(M_1) + \mathrm{res}(M_2)
$$
we see that $\mathrm{res}(M_C) \le \sum_{g \in s} \mathrm{res}(M_g) = m(C)$.
Additionally, $\mathrm{res}(M_C)=0$ if and only if $M_C = I_{2n}$, that is, $C$ is a Pauli unitary.
This shows that $h(M_C) = \mathrm{res}(M_C)$ is an admissible heuristic.
Using a similar argument, one can show that $h(M_C)$ is a consistent heuristic.

\section{Conclusions}

Another notable result is Theorem~5.1 in~\cite{C1976}, which says that any symplectic matrix $M$ over $\f_2$ can be expressed as 
a product of either $\mathrm{res}(M)$ or $\mathrm{res}(M)+1$ involutions. That is, any Clifford unitary $C$ can be written as 
a product of $\mathrm{res}(M_C)$ or $\mathrm{res}(M_C)+1$ generalized $S$ gates, a Pauli unitary, and a global phase.
This immediately implies that an $n$-qubit unitary can be written using at most~$2n+1$ generalized $S$ gates. 
However, this result does not immediately imply anything about the commutativity of the corresponding measurements needed to implement the $S$
gates and so provides less flexibility in space-time trade-offs.
An algorithm that decomposes a Clifford using a minimal number of generalized $S$ gates is discussed in~\cite{PVT2021}.
The proof of Theorem~5.1 in~\cite{C1976} is quite a bit more involved, in comparison to the simple constructive Theorem~4 in~\cite{YR2023}, which 
shows that $3n-3$ generalized $S$ gates are sufficient to implement any Clifford unitary.

As we noted before, $\mathrm{res}(M_C)$ provides a lower bound on the number of Pauli measurements needed to synthesize a Clifford unitary using 
generalized $S$ and $CZ$ gates. It is an open question whether this lower bound is tight. One way to address this would be to 
synthesize exceptional symplectic matrices using the A* algorithm outlined in \cref{sec:a-star}.
Similarly, it is unknown whether we can synthesize a Clifford unitary using two sets of mutually-commuting Pauli measurements and use only
$\mathrm{res}(M_C)$ of them.

\section*{Acknowledgments}
This work was completed while V.K. was a researcher at Microsoft Quantum.
 
\bibliographystyle{plainurl}
\bibliography{references}

\end{document}